\documentstyle[preprint,aps,prl,floats,epsf]{revtex}
\tighten
\begin{document}
\draft

\preprint{TIFR/TH/99-27}
\title{Production and Equilibration of the Quark-Gluon Plasma\\ 
with Chromoelectric Field and Minijets}

\author{R.S. Bhalerao \thanks{e-mail: bhalerao@theory.tifr.res.in}
and Gouranga C. Nayak \thanks{e-mail: nayak@th.physik.uni-frankfurt.de}
\thanks{present address: I.T.P., J.W. Goethe-Univ. Frankfurt, Germany}}

\address{Department of Theoretical Physics, Tata Institute of 
Fundamental Research, Homi Bhabha Road, Colaba, Mumbai 400 005, INDIA}

\maketitle

\begin{abstract}

Production and equilibration of quark-gluon plasma are studied within
the color flux-tube model, at the RHIC and LHC energies. Non-Abelian
relativistic transport equations for quarks, antiquarks and gluons,
are solved in the extended phase space which includes coordinates,
momenta and color. Before the chromoelectric field is formed, hard and
semihard partons are produced via minijets which provide the initial
conditions necessary to solve the transport equations. The model
predicts that in spite of the vast difference between the RHIC and LHC
incident energies, once the local equilibrium is reached, the energy
densities, the number densities and the temperatures at the two
machines may not be very different from each other. The minijet input
significantly alters the evolution of the deconfined matter, unless
the color field is too strong. For the input parameters used here the
equilibration time is estimated to be $\sim 1$ fm at RHIC and $\sim
0.5$ fm at LHC, measured from the instant when the two colliding
nuclei have just passed through each other. The temperature at
equilibration is found to be $\sim 250$ MeV at RHIC and $\sim 300$ MeV
at LHC.

\end{abstract}

\vspace{0.5 cm}


\section{INTRODUCTION}

Lattice QCD studies indicate that the hadronic matter undergoes a
phase transition to the deconfined quark-gluon plasma (QGP) phase, at
a sufficiently high temperature ($\simeq$ 200 MeV) \cite{LGT}. The
energy density at which this transition is expected to occur is about
2 GeV/fm$^3$. Such a phase of deconfined partons did exist in the very
early universe and is expected to be created in the laboratory, in
ultrarelativistic heavy-ion collisions (URHIC). The Relativistic
Heavy-Ion Collider (RHIC) at BNL (Au-Au collisions at $(\sqrt{s})^{NN}
=$ 200 GeV) and the Large Hadron Collider (LHC) at CERN (Pb-Pb
collisions at $(\sqrt{s})^{NN} =$ 5.5 TeV) will provide the best
opportunity to study the formation of QGP. The main signatures of
formation of QGP are sought in (1) $J/{\psi}$ suppression, (2)
strangeness enhancement and (3) dilepton and photon
productions \cite{qm99}.

Different stages through which the evolution of deconfined quark-gluon
matter proceeds are (1) pre-equilibrium (2) equilibrium and (3)
hadronization. While the equilibrium stage is described by the well
known hydrodynamic equations \cite{Bj83}, it is a difficult problem to
study the formation and pre-equilibrium evolution of QGP in URHIC.
Nevertheless, the initial conditions which one assumes for the
hydrodynamic expansion, can only be determined by a careful study of
the pre-equilibrium stage. So it is necessary and important to study
the pre-equilibrium stage, which would provide information on the
equilibration time, and the bulk properties such as the temperature,
the energy density and the number density of QGP. For example, the
importance of the correct determination of the equilibration time, in
connection with the $J/{\psi}$ suppression, is stressed by many people
\cite{gcn1,gcn2,xu} and it is a challenging task to determine this
quantity accurately.

In this paper, we study the pre-equilibrium evolution of QGP within
the color flux-tube model.
In this model,
when two Lorentz-contracted nuclei collide head-on and pass through
each other, they acquire a nonzero color charge ($<Q> = 0, <Q^2> \neq
0$), by a random exchange of soft gluons. The two receding nuclei act
as color capacitor plates and produce a chromoelectric field between
them \cite{low,nussinov}. This chromoelectric field causes instability
of the QCD vacuum and creates $q\bar{q}$ and gluon pairs via the
Schwinger-like mechanism \cite{sch,casher,gyulassy}. The partons so
produced, collide with each other and also get accelerated due to the
background field. These color charges also rotate in the color space
according to Wong's equation \cite{wong}. We consider
relativistic non-Abelian transport equations for quarks, antiquarks
and gluons. These equations simultaneously include the convective flow
terms, the background field acceleration terms, the $q\bar{q}$ and
gluon production source terms, the collision terms and the terms which
take into account rotations of these color charges due to the
background chromoelectric field. We solve these transport equations
together with the Yang-Mills equations.

As mentioned above, the initial chromoelectric field owes its
existence to nonperturbative processes, viz., the soft-gluon exchanges
that take place between the two nuclei during the collision.  This
model describes mostly the production of soft partons
\cite{nayak1,nayak2}. As pointed out in \cite{eskola} and also in
\cite{nayak2}, the (semi)hard processes (via perturbative QCD) take
place before the soft-gluon exchanges. By adding these hard processes
to the color flux-tube model, {\it we attempt to cover the full range
of particle production mechanisms, thereby including both soft and
hard partons in the evolution of the deconfined parton matter.} It has
been shown by several authors that minijet production is dominant in
the very early stages of the collision of two nuclei at the RHIC and
LHC energies \cite{kajantie}. 
Earlier studies
\cite{bbr,nayak1,nayak2} of the pre-equilibrium domain ignored the
minijets.
In this paper, we have calculated the
minijet production at RHIC and LHC and have combined it with the color
flux-tube model to study the production and evolution of QGP. A
similar attempt was made by Eskola and Gyulassy \cite{eskola} within
the hydrodynamics framework which is applicable only {\it after} local
equilibrium is attained. Here we study the {\it pre-equilibrium}
evolution by solving relativistic non-Abelian transport equations with
both the hard and soft partons taken into account. 
{\it We address ourselves to the important question whether
QGP, if formed, would attain substantially higher temperatures at LHC
than at RHIC.} We also investigate relative importance of the hard and
soft production mechanisms.

The plan of the paper is as follows. In section II, we describe our
model with a detailed discussion of minijet production and of
transport equations with their analytic solutions. Here we also
present our computational procedure. In section III, we describe our
numerical results and discuss them. We then compare our model with
some other models of the early evolution of the matter in URHIC.
Finally in section IV, we present
our conclusions.

\section{MODEL}

In our model, all hard and semihard partons (i.e., minijets) are
formed before the chromoelectric field is created due to the exchange
of soft gluons. In all earlier studies of pre-equilibrium evolution of
QGP within the color flux-tube model \cite{bbr,nayak1,nayak2}, there
was only the background field and no particles, at the initial
time. In other words, the initial particle distribution functions were
taken to be zero in all these calculations. In the present work, we
obtain the initial particle distributions from minijets. With this
initial condition, we then solve the transport equations at the RHIC
and LHC energies. Our aim is to study the possible effect of the
minijets on the equilibration process.

The relativistic non-Abelian transport equations which we want to
solve are presented in \cite{heinz,nayak2}, in detail.
In order to include the color charge in the
phase space, one considers an extended one-particle phase space of
dimension $d = 6 + (N^2 - 1)$, (with $N=3$). The extended phase space
is taken to be the direct sum $R^6 \oplus G$, where $G$ is the
(compact) space corresponding to the given gauge group. Thus, in
addition to the usual 3 space coordinates and 3 momentum coordinates,
one now has 8 coordinates corresponding to the 8 color charges in
SU(3). In this extended phase space, a typical transport equation
in the notation of \cite{heinz}, reads as 
\begin{equation}
\left[ p_{\mu} \partial^\mu + g Q^a F_{\mu\nu}^a p^\nu 
\partial^\mu_p
+ g f^{abc} Q^a A^b_\mu
p^{\mu} \partial_Q^c \right]  f(x,p,Q)=C(x,p,Q)+S(x,p,Q).
\label{trans}
\end{equation}
Here $f(x,p,Q)$ is the single-particle distribution function in the
extended phase space. The first term on the LHS of Eq.\ (\ref{trans})
corresponds to the usual convective flow, the second term is the
non-Abelian version of the Lorentz force term and the third term
corresponds to the precession of the color charge, as described by
Wong's equation \cite{wong}. $C$ and $S$ on the RHS of Eq.\
(\ref{trans}) are the collision and the source terms, respectively.
Note that there are separate transport equations for quarks,
antiquarks and gluons \cite{heinz}.

In order to solve the transport equations, we need the initial
distribution functions $f_0$, which we now obtain using minijet
production cross sections.

\subsection{Minijet production at RHIC and LHC}

Hard and semihard partons expected to be produced at RHIC and LHC are
mostly minijets whose average momenta are not very large. One can
calculate the minijet cross sections after fixing a momentum scale
($p_T$) above which perturbative QCD (pQCD) is applicable. In the
lowest order pQCD the inclusive ($2 \rightarrow 2$) minijet cross
section per nucleon in AA collision is given by \cite{eskola,kajantie}
\begin{equation}
\sigma_{jet} = \int dp_T dy_1 dy_2 {{2 \pi p_T} \over {\hat{s}}} 
\sum_{ijkl}
x_1~ f_{i/A}(x_1, p_T^2)~ x_2~ f_{j/A}(x_2, p_T^2)~
\hat{\sigma}_{ij \rightarrow kl}(\hat{s}, \hat{t}, \hat{u}).
\label{jet}
\end{equation}
Here $x_1$ and $x_2$ are the light-cone momentum fractions carried by
the partons $i$ and $j$ from the projectile and the target,
respectively, $f$ are the bound-nucleon structure functions and $y_1$
and $y_2$ are the rapidities of the scattered partons. The symbols
with carets refer to the parton-parton c.m. system. The
$\hat{\sigma}_{ij \rightarrow kl}$ is the elementary pQCD parton cross
section. The required $\hat{\sigma}_{ij \rightarrow kl}$ are given by
\begin{equation}
\hat{\sigma}_{q q^{\prime} \rightarrow q q^{\prime}} = 
{{4 \alpha_s^2} \over
{9 \hat{s}}} [{{\hat{s}^2+\hat{u}^2} \over {\hat{t}^2}}], \hspace{0.5cm}
\nonumber
\end{equation}
\begin{equation}
\hat{\sigma}_{q \bar{q} \rightarrow {q' \bar{q}'}} = {{4 \alpha_s^2} 
\over
{9 \hat{s}}} [{{\hat{t}^2+\hat{u}^2} \over {\hat{s}^2}}], \hspace{0.5cm}
\nonumber
\end{equation}
\begin{equation}
\hat{\sigma}_{q q \rightarrow q q} = {{4 \alpha_s^2} \over
{9 \hat{s}}} [{{\hat{s}^2+\hat{u}^2} \over {\hat{t}^2}} +
{{\hat{s}^2+\hat{t}^2} \over {\hat{u}^2}} -
{{2\hat{s}^2} \over {3\hat{t}\hat{u}}}], \nonumber
\nonumber
\end{equation}
\begin{equation}
\hat{\sigma}_{q \bar{q} \rightarrow {q \bar{q}}} = {{4 \alpha_s^2} 
\over {9 \hat{s}}} [{{\hat{s}^2+\hat{u}^2} \over {\hat{t}^2}} +
{{\hat{t}^2+\hat{u}^2} \over {\hat{s}^2}} -
{{2\hat{u}^2} \over {3\hat{s}\hat{t}}}], \hspace{0.5cm}
\nonumber
\end{equation}
\begin{equation}
\hat{\sigma}_{q \bar{q} \rightarrow {gg}} = {{8 \alpha_s^2} \over
{3 \hat{s}}} ({\hat{t}^2+\hat{u}^2}) [ 
{{4} \over {9\hat{t}\hat{u}}} - {{1} \over {\hat{s}^2}}], \hspace{0.5cm}
\end{equation}
\begin{equation}
\hat{\sigma}_{gg \rightarrow q \bar{q}} = {{3 \alpha_s^2} \over
{8 \hat{s}}} ({\hat{t}^2+\hat{u}^2}) [ 
{{4} \over {9\hat{t}\hat{u}}} - {{1} \over {\hat{s}^2}}], \hspace{0.5cm}
\nonumber
\end{equation}
\begin{equation}
\hat{\sigma}_{gq \rightarrow gq} = {{\alpha_s^2} \over
{ \hat{s}}} ({\hat{s}^2+\hat{u}^2}) [ 
{{1} \over {\hat{t}^2}}
- {{4} \over {9\hat{s}\hat{u}}}], 
\nonumber
\end{equation}
\begin{equation}
\hat{\sigma}_{gg \rightarrow gg} = {{9 \alpha_s^2} \over
{2 \hat{s}}} [ 3 - {{\hat{u}\hat{t}} \over {\hat{s}^2}}
 - {{\hat{u}\hat{s}} \over {\hat{t}^2}}
 - {{\hat{s}\hat{t}} \over {\hat{u}^2}}].
\nonumber
\end{equation}
Here $\alpha_s$ is the strong coupling constant, $q$ and $q'$ denote
distinct flavors of quark, and
\begin{equation}
\hat{s}=x_1 x_2 s = 4 p_T^2 ~{\rm cosh}^2
\left ( {{y_1-y_2} \over {2}} \right ).
\nonumber 
\end{equation}
The rapidities $y_1$, $y_2$ and the momentum fractions $x_1$, $x_2$ are
related by,
\begin{equation}
x_1=p_T~(e^{y_1}+e^{y_2})/{\sqrt{s}}, \hspace{0.5cm}
x_2=p_T~(e^{-y_1}+e^{-y_2})/{\sqrt{s}}.
\nonumber
\end{equation}
The limits of integrations of rapidities $y_1$ and $y_2$ are given by,
$\vert {y_1} \vert \leq$ ln($\sqrt{s}/{2 p_T} + \sqrt{s/{4 p_T^2}
-1}$) and $-$ln(${\sqrt{s}/{p_T}-e^{-y_1}}) \leq y_2 \leq$
ln(${\sqrt{s}/{p_T}-e^{y_1}})$. We multiply the above minijet cross
sections by the well known factor $K=2$, to account for the
$O(\alpha_s^3)$ contributions.

The structure functions $f_{i,j/A}$ occurring in Eq.\ (\ref{jet}) are
obtained from the model of Eskola {\it et al.} \cite{eskola1,eskola2},
and using the GRV HO 94 set of parton distributions for a free
nucleon.

As argued in \cite{eskola}, we choose minimum $p_T = 2$ GeV, above
which pQCD is assumed to be applicable.

\subsection{Initial conditions for plasma evolution from minijets}

For central collisions, the minijet cross section (Eq.\ (\ref{jet}))
can be related to the total number of partons $(N)$ by
\begin{equation}
N=T(0) ~\sigma_{jet},
\label{number}
\end{equation}
where $T(0)= 9A^2/{8\pi R_A^2}$ is the nuclear geometrical factor for
head-on AA collisions (for a nucleus with a sharp surface)
\cite{kajantie}. Here $R_A=1.2 A^{1/3}$ is the nuclear radius. The
total transverse energy $<E_T^{tot}>$ produced in nuclear collisions,
from minijets, can be shown to be
\begin{equation}
<E_T^{tot}> = T(0) \int dp_T p_T dy_1 dy_2 {{2 \pi p_T}
 \over {\hat{s}}} \sum_{ijkl} x_1 ~f_{i/A}(x_1, p_T^2) ~x_2 
 ~f_{j/A}(x_2, p_T^2)
 ~\hat{\sigma}_{ij \rightarrow kl}(\hat{s}, \hat{t}, \hat{u}).
\label{tenergy}
\end{equation}
We will use these relations (Eqs.\ (\ref{jet}), (\ref{number}) and
(\ref{tenergy})) to determine the initial conditions which are
required to solve the transport equations (Eq.\ (\ref{trans})) in the
pre-equilibrium stage.

In the color flux-tube model, we assume that the chromoelectric field
is formed when the two nuclei completely pass through each other. The
volume of the system at that instant is $V_0 = \pi R_A^2 d$, where $d$
is the longitudinal thickness of the system. Note that $d$ depends on
the incident energy because of the Lorentz contraction and the
presence of the wee partons in the incoming nuclei. We have taken $d =
3$ fm at RHIC for Au-Au collisions and $d = 2$ fm at LHC for Pb-Pb
collisions \cite{kg95}. Using (Eqs.\ (\ref{jet}), (\ref{number}) and
(\ref{tenergy})), we calculate the initial number density ($n_0$),
initial energy density ($\epsilon_0$) and initial distribution
functions $(f_0)$ of quarks, antiquarks and gluons, which are needed
to solve the transport equations. These are $n_0=N/{V_0}$,
$\epsilon_0=<E_T^{tot}>/{V_0}$ and $f_0=dn_0/d^3p$. Here
\begin{equation}
d^3p=d^2p_T dp_z=p_T ~d^2p_T ~{\rm cosh} {y_1} ~{dy_1},
\label{d3p}
\end{equation}
and $f_0$ can be easily extracted from Eq.\ (\ref {number}) with the
help of Eqs.\ (\ref{jet}) and (\ref{d3p}).

In Table I we present our results for the initial conditions, obtained
from minijet production at RHIC and LHC.

\subsection{Solution of the relativistic transport equations}

In order to solve the transport equations (\ref {trans}), we need to
specify the collision and the source terms occurring on the RHS. 

We measure the proper time $\tau$ from the instant the two colliding
nuclei have just passed through each other. In our model, hard
partons (in the form of the minijets) are present in the system at
$\tau = 0$ and soft partons are created by the background field at
later times. It is not possible to obtain the collision term from pQCD
alone because the latter provides a framework to treat only the {\it
hard} collisions; collisions among {\it soft} partons cannot be
handled by pQCD. Including the effects of soft collisions, in the
transport equations (1), from first principles (QCD) is a difficult
problem. We make the relaxation-time approximation to simulate the
effects of both soft and hard collisions and write the collision term
as:
\begin{equation}
C = -p^\mu u_\mu (f-f^{eq}) / {\tau_c},
\end{equation}
where $u^\mu$ is the four velocity, $f^{eq}$ is the equilibrium
distribution function (Fermi-Dirac for quarks and antiquarks, and
Bose-Einstein for gluons) and $\tau_c$ is the relaxation time. We take
the same expressions for the source terms ($S$) as in \cite{nayak2}.

Following Bjorken's proposal \cite{Bj83}, we express all physical
observables in terms of the boost invariant quantities, namely
\begin{equation}
\tau =(t^2-z^2)^{1/2}, ~~\xi=\eta-y, ~~p_T=(p_0^2-p_l^2)^{1/2},
\end{equation}
where $\eta=\tanh^{-1}(z/t)$ is the space-time rapidity and
$y=\tanh^{-1}(p_l/p_0)$ is the momentum rapidity. The transport
equations (\ref{trans}) can be rewritten in terms of these variables.

The formal solutions of the transport equations can be found to be,
\begin{eqnarray}
f_{q,{\bar q},g}(\tau,\xi,p_T,Q) &=& \int^\tau_0 d \tau^\prime 
\exp \left ( \frac{\tau^\prime-
\tau}{\tau_c} \right ) \left[ \frac{S_{q,{\bar q},g}
(\tau^\prime,\xi^\prime,p_T,Q )}
{p_T \cosh \xi^\prime} + \frac{f_{q,{\bar q},g}^{eq}
(\tau^\prime,\xi^\prime,p_T,Q)}
{\tau_c} \right] \nonumber \\ \nonumber \\
& & +~ \exp (-\tau/{\tau_c}) f^{q,{\bar q},g}_0,
\label{dist}
\end{eqnarray}
where $\xi^{\prime}$ is given by
\begin{equation}
\xi^\prime= \sinh^{-1} \left[ \frac{\tau}{\tau^\prime}
\sinh \xi
+ \frac{g\cos\theta_1}{p_T \tau^\prime} 
\int_{\tau^\prime}^ {\tau} 
d\tau^{\prime\prime}~\tau^{\prime\prime}~E(\tau^{\prime\prime}) \right],
\end{equation}
and $f^{q,{\bar q},g}_0$ are the initial distribution functions of
quarks, antiquarks and gluons obtained from minijets (as described in
subsections IIA-B). The initial distribution function was assumed to
be zero in all earlier studies of the color flux-tube model
\cite{bbr,nayak1,nayak2}. In the above equation, $\theta_1$ is the
angle between the chromoelectric field and the color charge in the
SU(3) color space \cite{nayak2}.

Since both the field and the particles are present simultaneously, we
use the following relation for the conservation of the energy-momentum
tensor:
\begin{equation}
\partial_\mu T^{\mu\nu}_{mat} + 
\partial_\mu T^{\mu\nu}_{f}=0.
\end{equation}
Here
\begin{equation}
T^{\mu\nu}_{mat}=\int p^\mu p^\nu 
(3 f_q + 3 f_{\bar q} + f_g)d\Gamma d {\Omega},
\label{tmat}
\end{equation}
and
\begin {equation}
T^{\mu\nu}_f = \mbox{diag} (E^2/2; E^2/2, E^2/2, -E^2/2 ).
\end{equation}
In Eq.\ (\ref{tmat}), the factor 3 refers to the three flavors of
quarks, $d \Gamma = d^3 p/{(2 \pi)^3 p_0} = p_T d p_T d \xi/{(2
\pi)^2}$, and $d \Omega$ is the angular integral measure in the color
space.
We follow the same procedure as in \cite{nayak2,bbr} and finally
obtain the evolution equation for the field as:
\begin{equation}
{ dE(\tau) \over d\tau }- \frac{g }{(2\pi)^2} 
\int^\infty_0 dp_T p_T^2
\int^\infty_{-\infty} d\xi\sinh\xi \int d\Omega
[3 f_q - 3 f_{\bar q} +f_g ] 
+({{\pi}^3/6})\bar{a} \vert E(\tau) \vert^{3/2} = 0.
\label{diff}
\end{equation}
Here $\bar{a} = a\zeta (5/2)\exp (0.25/\alpha )$,  
$ a = c(g/2)^{5/2}\frac{9}{2 (2\pi)^3}$,  
$ c = \frac{2.876}{4\pi^3}$
and $\zeta (5/2) =1.342$ is the Riemann zeta function.

To solve Eq.\ (\ref{diff}), we fix the form of the local temperature
$T(\tau)$, by demanding that the particle energy density
$\epsilon_p(\tau)$ at any instant, differs negligibly from the
equilibrium energy density at temperature $T(\tau)$ at that
instant. (It may be noted that $T(\tau)$ occurs in Eq.\ (\ref{dist})
and hence in Eq.\ (\ref{diff}), through $f^{eq}$.) This allows us to
write the temperature $T(\tau)$ in terms of the particle energy
density $\epsilon_p(\tau)$ as follows
\begin{equation}
T(\tau)= \left [ {{36 \epsilon_p(\tau)} \over {5 \pi^6}} \right ] ^{1/4},
\label{temp}
\end{equation}
where
\begin{equation}
\epsilon_p(\tau)=\int d\Gamma d {\Omega} (p^\mu u_\mu)^2 
(3 f_q + 3 f_{\bar q} +f_g).
\label{energy}
\end{equation}
We solve Eq.\ (\ref{diff}) numerically to study time evolution of the
chromoelectric field.

Finally, the number density of the quark-gluon matter is given by
\begin{equation}
n(\tau)=\int d\Gamma d {\Omega} p^\mu u_\mu
(3 f_q + 3 f_{\bar q} +f_g).
\label{ntau}
\end{equation}

\subsection{Computational procedure}

In the present work, we have determined $E(\tau)$ and $T(\tau)$ by
imposing the following double self-consistency requirement. Starting
with trial $E(\tau)$ and $T(\tau)$ and using Eqs. (\ref{dist}) we
determine $f_q$, $f_{\bar q}$ and $f_g$. These when substituted in
Eq. (\ref{energy}) give us $\epsilon_p(\tau)$ which in turn gives a
new temperature $T(\tau)$ by Eq. (\ref{temp}). The new $T(\tau)$ and
the trial $E(\tau)$ are again used in Eqs. (\ref{dist}), and with the
new $f_q$, $f_{\bar q}$ and $f_g$ thus determined, the differential
equation (\ref{diff}) is solved to get a new field $E(\tau)$. This
process is iterated until convergence is reached. This gives us a
self-consistent set of $E(\tau)$, $T(\tau)$, $\epsilon_p(\tau)$,
$f_q$, $f_{\bar q}$ and $f_g$.

\section{RESULTS AND DISCUSSION}

With the minijet initial conditions, we solve the transport equations
(1) from the instant the two wounded nuclei have just crossed each
other up to $\tau \simeq 1.5$ fm.
The solution of the transport equations allows us to determine the
time evolution of the physical quantities such as energy density,
number density and temperature. The determination of these quantities
is important for the prediction of any signature of quark-gluon
plasma. We now present our numerical results for these quantities. As
in \cite{nayak2}, we take $g=4$ and $\tau_c = 0.2$ fm. For the initial
field energy density $\epsilon_f(0)$, we take either $0$ or 300
GeV/fm$^3$. This allows us to discuss the following three scenarios:
(a) vanishing field, (b) $\epsilon_f(0)$ comparable to the initial
minijet energy density $\epsilon_0$, and (c) $\epsilon_f(0) \gg
\epsilon_0$. Thus we can study the effect of the variation of the
initial field on the calculated results.

In all the figures, $\tau = 0$ corresponds to the instant when the two
colliding nuclei have just passed through each other, and
$\epsilon_f(0)=300$ GeV/fm$^3$ unless stated otherwise.

In Fig. 1, we have presented the parton energy densities for RHIC and
LHC. It can be seen that the evolution of the parton energy density
with minijet inputs at RHIC, is almost the same as that without the
minijet inputs (dashed curve). This is because, at RHIC, the initial
minijet energy density $\epsilon_0$ ($\simeq 3$ GeV/fm$^3$) is
negligible compared to $\epsilon_f(0)$; see Table I. At LHC, the two
results are different because $\epsilon_0$ ($\simeq 178$ GeV/fm$^3$)
is comparable to $\epsilon_f(0)$. {\it Thus the inclusion of the
minijet input substantially alters the evolution of the parton energy
density unless the chromoelectric field is too strong.}

We now discuss qualitatively the shapes of the curves in Fig. 1.
It is easy to see from Eq.\ (\ref{dist}) that the parton energy
density evolves as
\begin{equation}
\epsilon_p(\tau)=\epsilon(\tau)+e^{- \tau/\tau_c}\epsilon_0.
\label{ept}
\end{equation} 
Here, $\epsilon(\tau)$ is the energy density if there is no minijet
production, i.e., if the particle production is only from the
field. The decrease in the energy density at earlier times (see the
curve labelled LHC in Fig. 1) is due to the multiplicative factor
$e^{- \tau/\tau_c}$ occurring in Eq.\ (\ref{ept}). The further
increase in the energy density is due to the fact that the particle
production from the field starts dominating by this time (this is
$\epsilon(\tau)$ in Eq.\ (\ref{ept})). We shall show below that the
fall in the energy density after it reaches the maximum value is close
to what is expected if the system has attained a local equilibrium and
is expanding according to the hydrodynamic equations. Then it is clear
from Fig. 1 that {\it the equilibration of plasma is expected to occur
earlier and the temperature at equilibration is expected to be
somewhat higher at LHC.} This will have a significant effect on all
the predictions of QGP signatures.

In Fig. 2, we compare the evolution of the parton energy density, at
LHC, with and without the background field. This is necessary because
the initial strength of the field cannot be calculated easily and as
the incident energy increases the possibility of soft-gluon exchanges
may become small. So we have presented our results for the two extreme
cases (1) vanishing field energy density and (2) a very high initial
field energy density (300 GeV/fm$^3$). The actual results are expected
to fall within these two limits. 
Qualitatively similar results are expected at the RHIC energy too.

Thus the background field representing the collective long-range
effects plays an important role in the evolution of the system. If the
minijet initial conditions based on pQCD are to play an important role
at RHIC, then the field will have to be about two orders of magnitude
weaker than what we have assumed. In other words, if at RHIC, the
initial field energy density $\epsilon_f(0)$ was not 300 GeV/fm$^3$,
but only 3 GeV/fm$^3$, then $\epsilon_f(0)$ would be comparable to the
initial minijet energy density $\epsilon_0$(RHIC) $\simeq 3$
GeV/fm$^3$, and results qualitatively similar to those shown in Fig. 2
would be obtained at the RHIC energy too. In the HIJING \cite{xnw97}
and Parton Cascade models \cite{kg95}, the background field was
completely neglected.

Comparison of the two curves labelled `Field' and `Field+Minijets', in
Fig (2), shows the effect of the minijets on $\epsilon_p(\tau)$. The
memory of the minijet input is nearly wiped out at large $\tau$
because of the exponential decay factor in Eq.\ (\ref{ept}). {\it As a
result, the large kinetic energy of the two colliding nuclei at LHC
may not necessarily translate into large temperature of the
equilibrated deconfined matter.} Actual calculation shows that the
temperature when the equilibrium is reached is $\sim 300$ MeV at LHC
and $\sim 250$ MeV at RHIC, {\it not very different from each
other}. Comparison of the two curves labelled `Minijets' and
`Field+Minijets' shows the effect of the field on $\epsilon_p(\tau)$;
this effect survives till relatively larger $\tau$.
Note also that {\it if there is no
background field, the $\epsilon_p(\tau)$ decreases monotonically.}

The decay of the chromoelectric field is shown in Fig. 3. The decay is
much faster at LHC than at RHIC, again suggesting an early
equilibration at LHC. The evolution of the field strength has a
significant impact on the acceleration of partons. If the
chromoelectric field survives even after local equilibrium is reached,
then the subsequent evolution of the system is governed by the
chromoviscous hydrodynamic equations \cite{eskola}, instead of
Bjorken's hydrodynamic equations \cite{Bj83}.  So it is very important
to study the evolution of the chromoelectric field in the
pre-equilibrium stage.
{\it We find that the equilibration time of the deconfined matter and
the life-time of the field are nearly the same.}

In Fig. 4, we have presented the parton number densities of the
system, for both the RHIC and LHC energies. The behavior of the
results is the same as that of the parton energy densities in Fig. 1,
because an equation similar to Eq.\ (\ref{ept}) holds for the number
density too. The maximum number density at LHC is $\sim 120$
fm$^{-3}$, whereas at RHIC it is $\sim 90$ fm$^{-3}$. 
{\it However, at large $\tau$, when the systems have equilibrated,  
the two number densities are nearly the same.} These
partons produce/affect all the signatures of QGP and hence their
evolution plays a crucial role in QGP detection.

We have fitted $\epsilon_p(\tau), ~n(\tau)$ and $T(\tau)$, for $\tau
\ge 0.5$ fm, at the LHC energy, with a functional form $a \tau ^
{-b}$, where $a$ and $b$ are constants. We find that $b = 1.23, ~
0.70$ and $0.31$, respectively. According to Bjorken's hydrodynamics,
these values are $1.33, ~1$ and $0.33$, respectively. Comparing these
two sets of values with each other, we conclude that the system has
nearly equilibrated at $\tau \simeq 0.5$ fm, at LHC. Similarly, at
RHIC, the equilibration time is found to be $\simeq 1$ fm. Recall that
these time intervals are measured from the instant when the two
colliding nuclei have just passed through each other.

In Figs. 5 and 6, we have presented our results for the number
densities of quarks plus antiquarks and gluons, respectively. Consider
the curves labelled LHC in these two figures. The initial number
density in Fig. 5 is much less than that in Fig. 6, because the
minijet gluon production is much larger; see Eqs. (3)-(10). In Fig. 5,
initially, the number density increases with time because of the large
production of $q$ and $\bar q$ from the field; see Eq.\ (\ref{ntau})
which receives a separate contribution from each of the 3 flavors of
quarks and antiquarks. Although the gluon production is 1.5 times more
likely than that of quarks via the Schwinger-like mechanism
\cite{gyulassy}, the counting in Eq.\ (\ref{ntau}) 
eventually builds up a larger
quark plus antiquark density than the gluon density. In Fig. 6, the
LHC number density decreases with time because of the dominance of the
exponential factor $ {\rm exp}(-\tau / \tau_c)$ in Eq.\ (\ref{dist}),
over the gluon production by the field. The energy densities for
quarks plus antiquarks and gluons are plotted in Figs. 7 and 8,
respectively. The behavior remains the same as in Figs. 5 and 6.

We now compare our model with some well known models of the early
stages of the ultrarelativistic heavy-ion collisions, namely the
parton cascade model (PCM) \cite{kg95}, the HIJING model \cite{xnw97},
the McLerran-Venugopalan model (MVM) \cite{lm94}, etc.

As is well known, PCM is a pQCD-based model. It neglects the
long-range collective effects which we have attempted to incorporate
here by means of a background field. The PCM starts by defining a
classical phase-space distribution function for the two incoming
nuclei, which is then evolved by means of the Boltzmann equation
retaining only the convective-flow term and the collision term in
Eq. (1). Now it is known that there are wee partons in each of the two
nuclei and each wee parton spreads over the entire width of the
nucleus \cite{bj75}. It is difficult to define a classical
distribution function for such a system of particles. Indeed the
incoming nuclei are in pure quantum mechanical states. Hence solving
the classical transport equation right from the instant the two
incoming nuclei start touching each other, is a questionable
procedure.

A classical distribution function can be more justifiably defined when
the spatial extent and the time intervals are larger than the average
de Broglie wave length of the partons. In our model, we have started
the classical evolution at a time when the two nuclei have completely
crossed each other. At that instant, in our model, there are only hard
partons ($p_T > 2$ GeV/c); the soft partons are created later when the
system has grown even bigger. The initial condition for the hard part
is obtained via the minijet calculation where no approximation is
involved. The initial condition for the soft part is taken through the
creation of a coherent chromoelectric field.

The HIJING model too is a pQCD-based model. It combines pQCD processes
with string phenomenology for non-perturbative soft processes. It {\it
assumes} that the parton distributions can be approximated by thermal
phase-space distributions with non-equilibrium fugacities. Thus the
issue of thermalization of the deconfined matter in the framework of
the kinetic theory is not addressed.

Thus we have presented a model which is different from PCM and the
HIJING Model. We believe that this model has some desirable features.
For simplicity, we have neglected the interactions among the minijets
before $\tau = 0$. Our main aim has been to see how the minijet input
modifies our earlier analysis \cite{nayak2}. It will be interesting to
consider these interactions. Although pQCD provides a general
framework for this purpose, including these interactions is quite
nontrivial.

Our approach has some similarities with the McLerran-Venugopalan
model. The MVM is a classical effective field theory description of
the small-$x$ modes in very large nuclei at very high energies. This
effective theory contains a scale $\mu$ which is proportional to the
large gluon density at small $x$. The large gluon density ensures that
even if the coupling is weak, the fields may be highly
nonperturbative. They argued that the classical fields corresponding
to the saddle point solutions of the effective theory are the
non-Abelian analogue of the Weizs\"acker-Williams fields in classical
electrodynamics. Now in a collision of such Weizs\"acker-Williams
fields, the production of incoherent partons with transverse momenta
$p_T >> \alpha_s \mu$ can be handled by means of pQCD and their
propagation in the coherent field can be studied with the help of a
classical transport equation \cite{lm99}. In our model too both
incoherent hard partons and coherent field (soft part) are taken into
account. So in this sense our approach is similar to MVM.

Real-time simulation of the full, nonperturbative evolution of the
classical non-Abelian Weizs\"acker-Williams fields on lattice, for the
gauge group $SU(2)$, has recently been studied by Krasnitz and
Venugopalan \cite {kv99}; see also \cite{wp99}. The calculation in
\cite {kv99} incorporates coherence effects which become important at
small $x$ and small $p_T$, while reproducing simultaneously the
standard minijet results at large $p_T$. However, the important issue
of equilibration of gluons was not addressed. It will also be
interesting to see both the coherent field and incoherent partons
simultaneously taken into account in such a simulation.

\section{CONCLUSIONS AND FUTURE PROSPECTS}

We have studied the production and equilibration of the quark-gluon
plasma expected to be formed in ultrarelativistic heavy-ion
collisions, at RHIC and LHC, within the color flux-tube model. The
distribution functions of quarks, antiquarks and gluons are defined in
the extended phase space of dimension 14 in SU(3). We have solved the
non-Abelian relativistic transport equations for these distribution
functions. The quarks, antiquarks and gluons are produced from the
background field which is formed due to soft-gluon exchanges, via the
Schwinger-like mechanism. The initial distribution functions of
partons are calculated from the minijet production at RHIC and LHC.

The model predicts that 

\noindent
({\it i}) The inclusion of the minijet input substantially alters the
evolution of the 
bulk properties of the deconfined matter unless the chromoelectric
field is too strong.

\noindent
({\it ii}) Equilibration of the deconfined matter is expected to be
faster at LHC than at RHIC. Estimated equilibration times are given. 

\noindent
({\it iii}) 
Surprisingly, in spite of the large difference between the LHC and
RHIC colliding beam energies, the differences between the temperatures
and equilibrium energy and number densities attained at the two
machines, may not be very large. This is because the memory of the
minijet input decays exponentially with time. Estimated numerical
values for these physical quantities are given.

\noindent
({\it iv}) Unlike the minijets, the effect of the background field
survives over a longer time.

\noindent
({\it v}) In the absence of a background field, the parton energy
density and the number density decrease monotonically with time.

\noindent
({\it vi}) The equilibration time of the deconfined matter and the 
life-time of the chromoelectric field are nearly the same.

In summary, we have combined the non-perturbative aspects of QGP
(background chromoelectric field formation) with the perturbative
aspects (minijet production) and have studied the evolution of the
quark-gluon plasma.

We have used in this work the Schwinger-like source terms
\cite{sch,casher,gyulassy} for parton production from a chromoelectric
field. This mechanism is strictly applicable only for a constant,
uniform field. However, in a heavy-ion collision, as the system
evolves, the field acquires space-time dependence. Particle production
in such a field has been studied and an appropriate source term has
been derived in \cite{bhal}, which will be incorporated in the
pre-equilibrium evolution of the QGP and results will be reported
elsewhere.

\bigskip\bigskip\bigskip

Note added:

Recently Bloch et al. have studied the thermodynamics of strong-field
plasmas \cite{jcrb99}. In the present paper, we used a classical
transport equation while they have used a quantum equation. Our
equation is non-Abelian, theirs is Abelian. Our initial conditions are
different from theirs. Our source term is also different from
theirs. But still the time dependence of the various quantities (their
Figs. 1-3) is qualitatively similar to that obtained by us. This shows
the robustness of our findings.

\acknowledgments

We thank R.V. Gavai for many useful discussions.

\newpage


\begin{table}[tb]
\caption{Initial conditions for pre-equilibrium evolution of QGP,
obtained from the minijet calculation.}
\medskip
\begin{tabular}{lccccc}
& $(\sqrt{s})^{NN}$ (GeV) & $N$ & $V_0$ (fm$^3$) & $n_0$ (fm$^{-3}$) 
& $\epsilon_0$ (GeV/fm$^3$) \\ \hline
RHIC & 200 & 504 & 459 & 1.1 & 3 \\ \hline
LHC & 5500 & 17422 & 318 & 55 & 178 \\
\end{tabular}
\end{table}

\newpage

\subsection*{Figure captions}

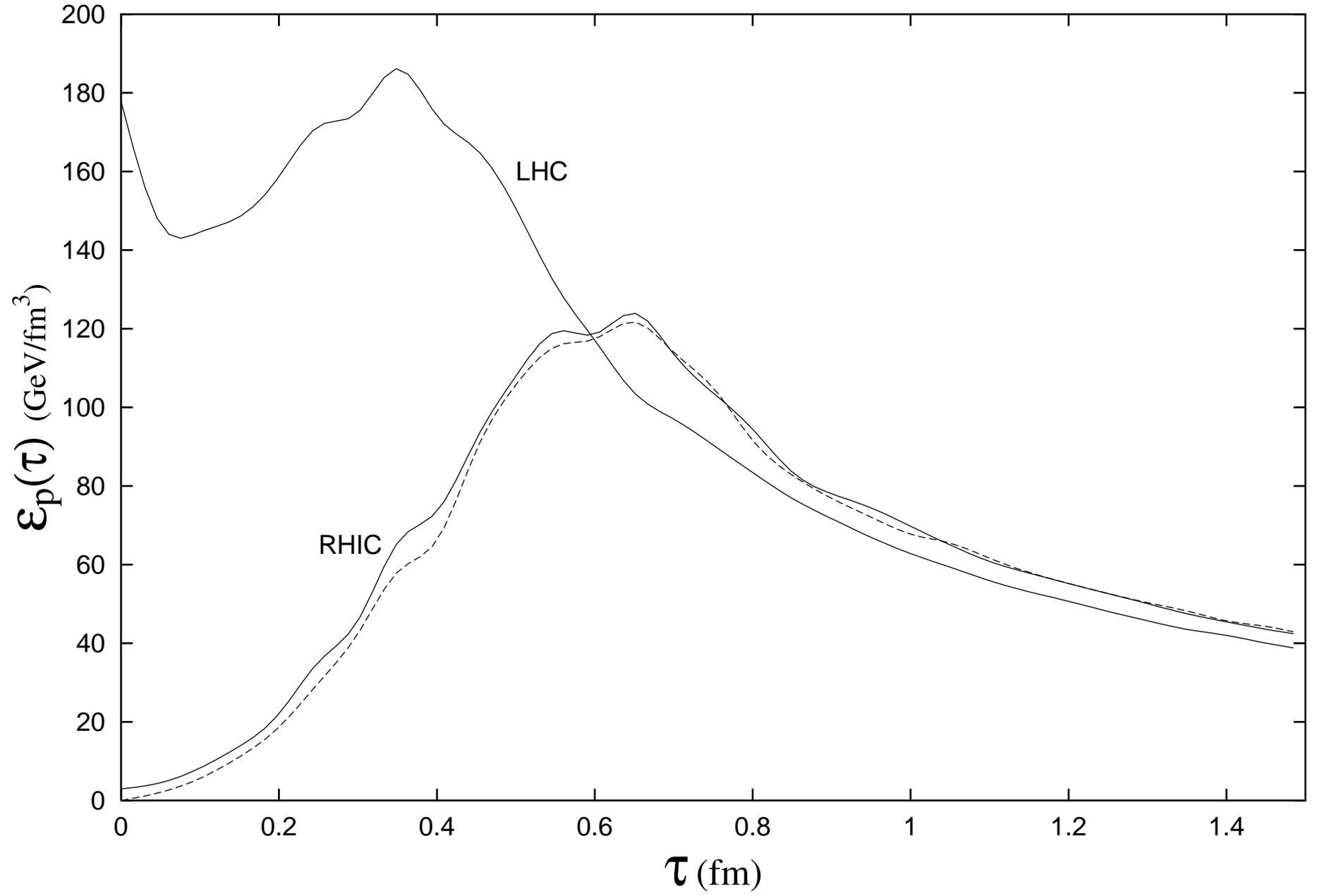
FIG. 1. Parton energy density vs proper time. The dashed line refers
to the calculation without the minijets.

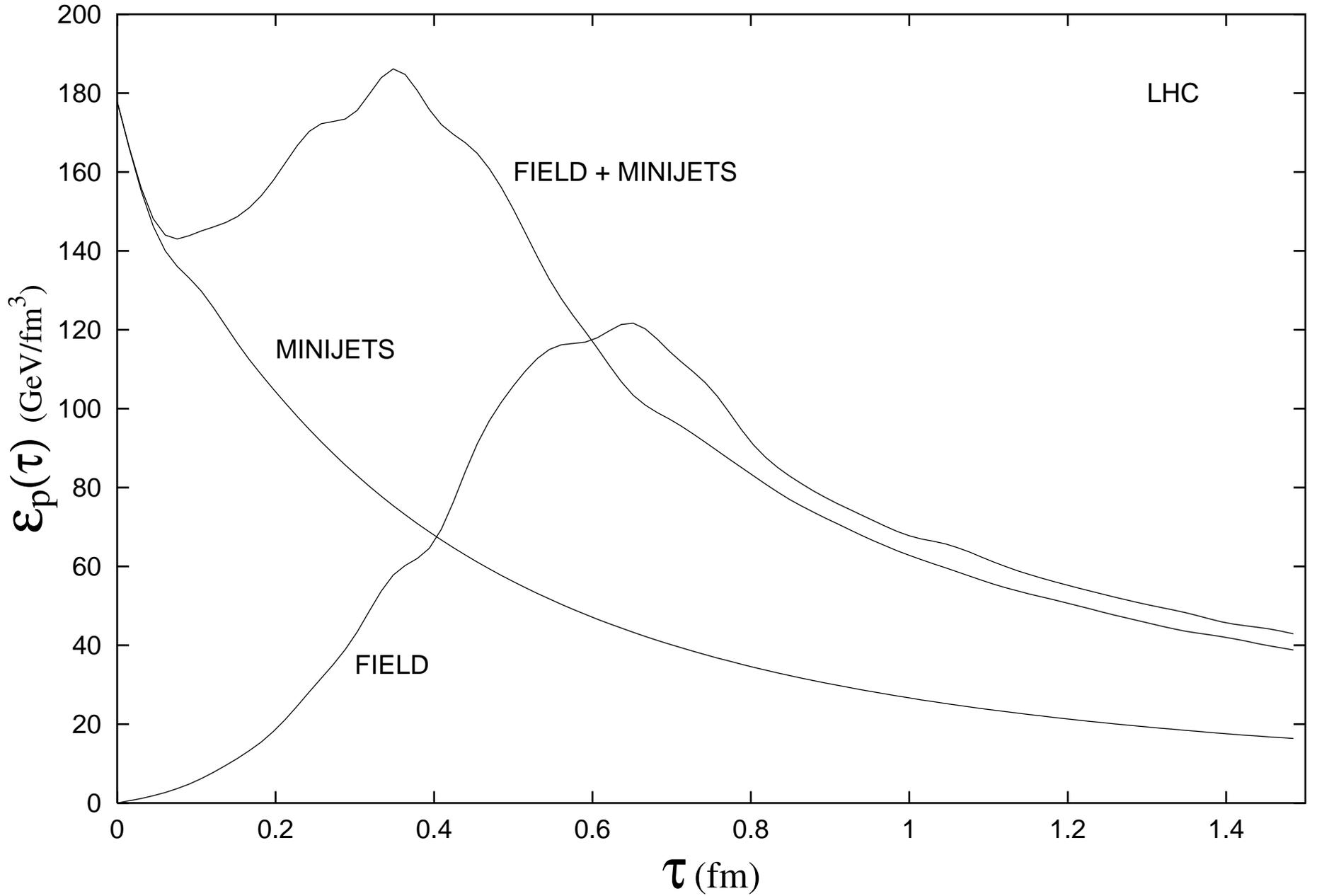
FIG. 2. Parton energy density vs proper time, at the LHC energy. The
curve labelled Field (Minijets) is based on the calculation without
the minijets (background field). The curve labelled Field + Minijets
is based on the full calculation with both the minijets and the
background field taken into account.

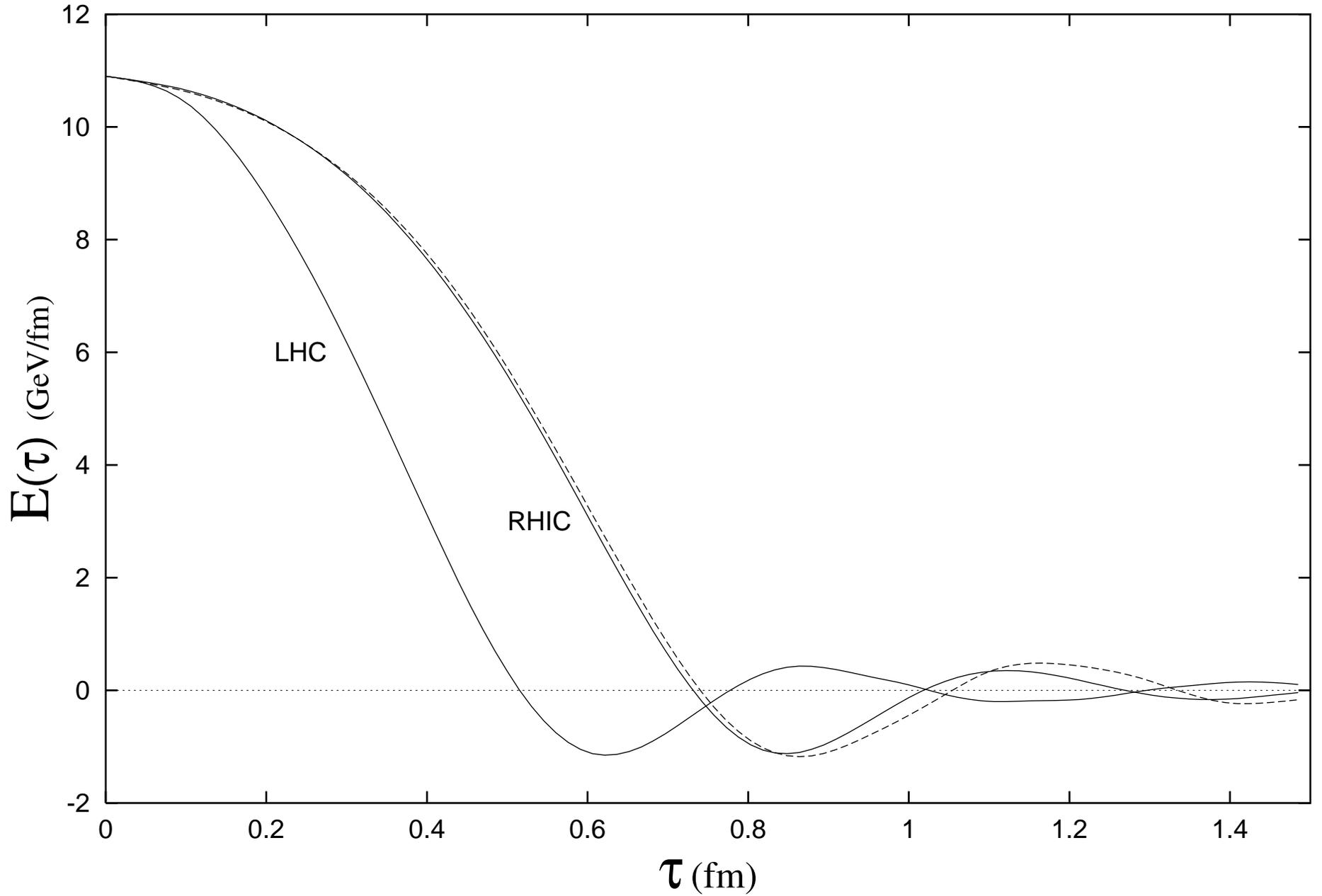
FIG. 3. Chromoelectric field vs proper time. The dashed line as in
Fig. 1.

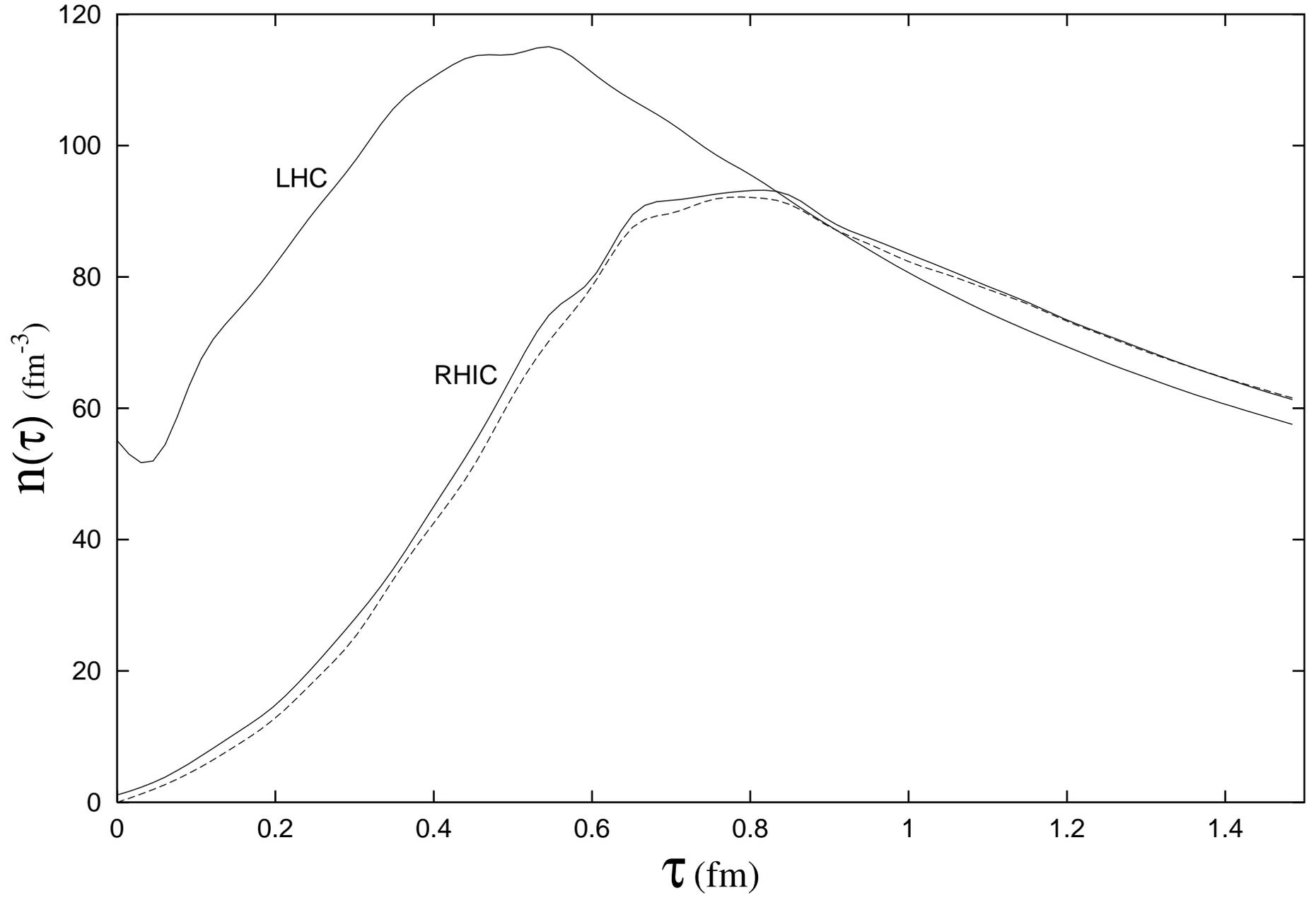
FIG. 4. Number density vs proper time. The dashed line as in Fig. 1.

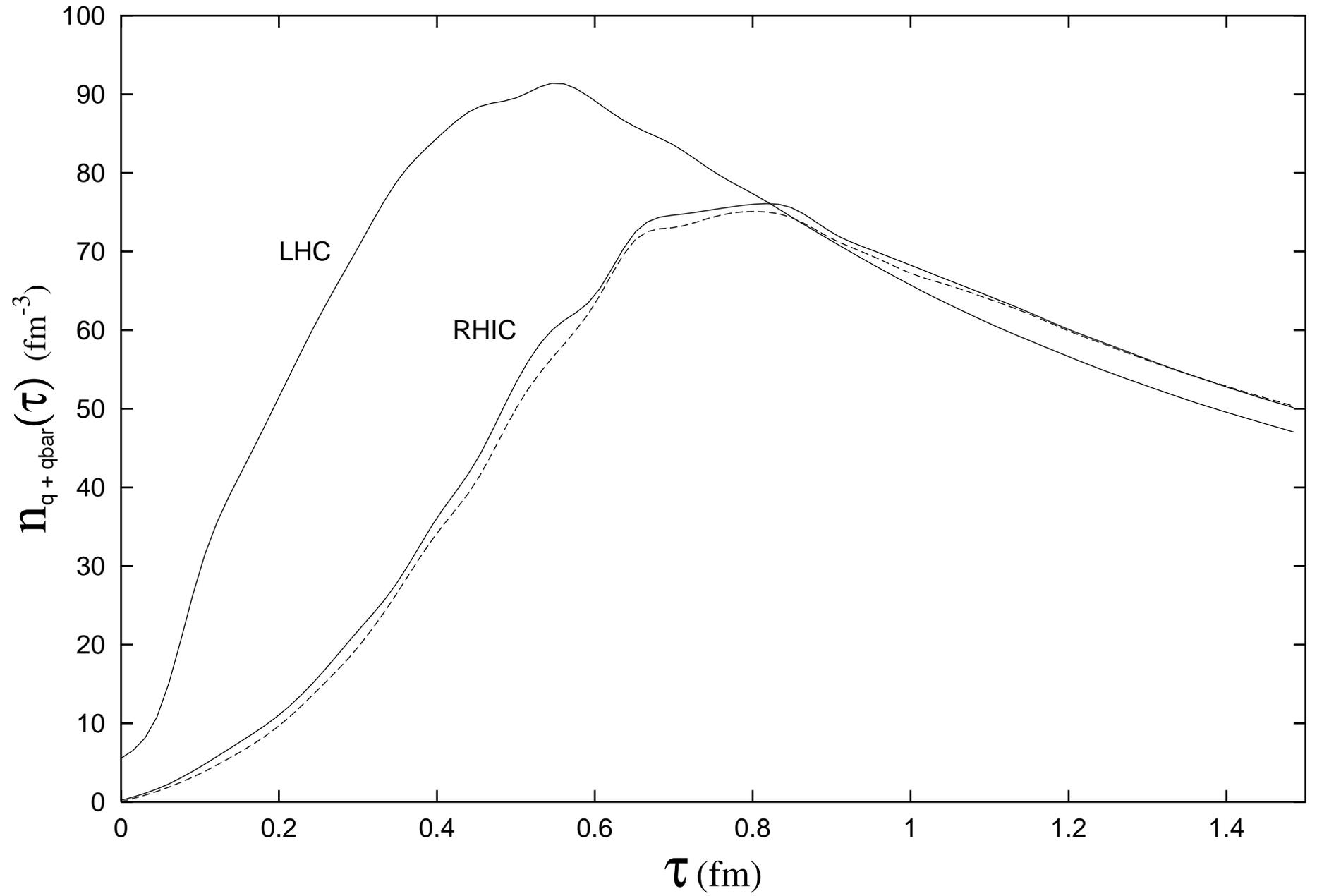
FIG. 5. Number density for quarks and antiquarks vs proper time. The
dashed line as in Fig. 1.

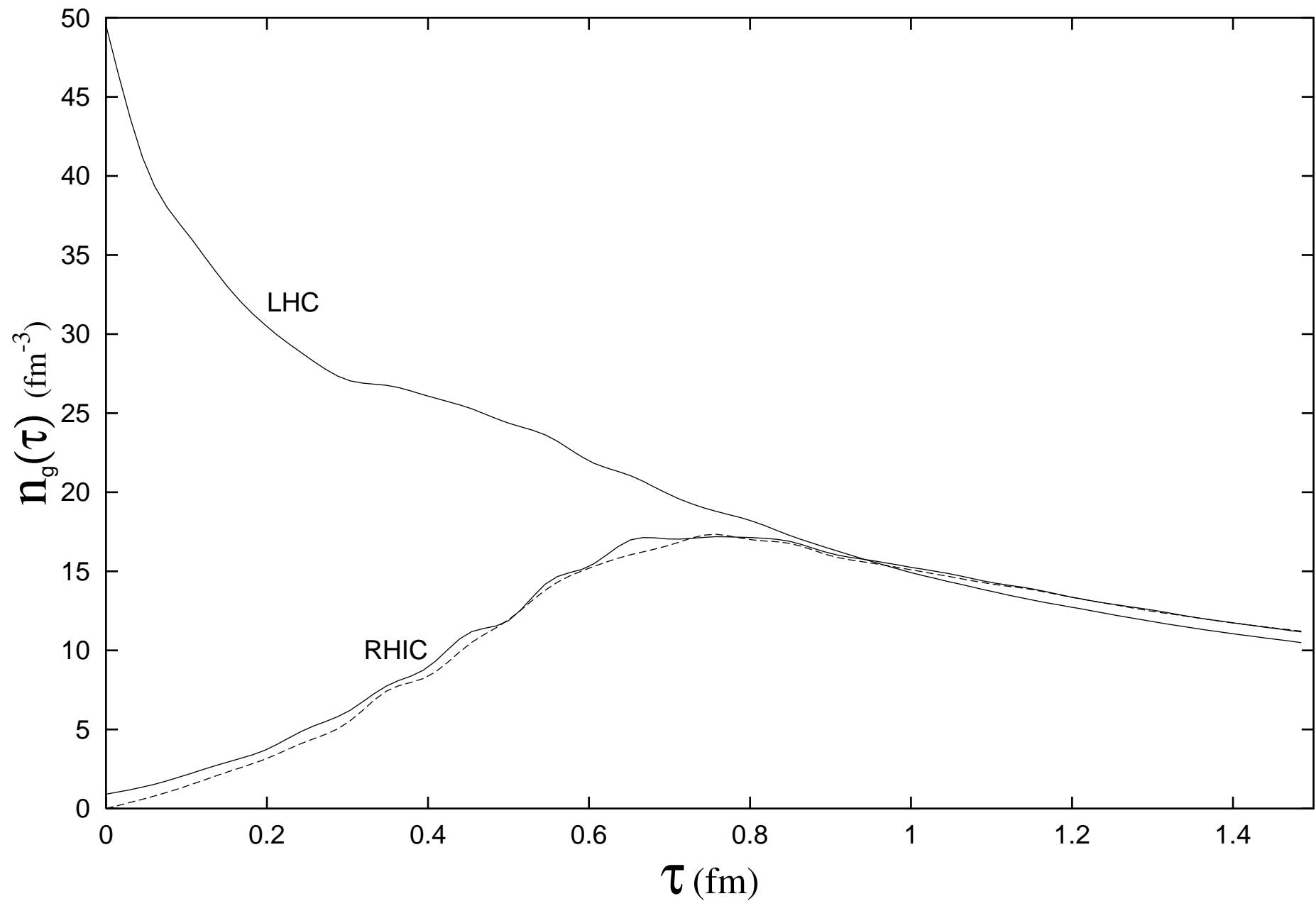
FIG. 6. Number density for gluons vs proper time. The dashed line as
in Fig. 1.

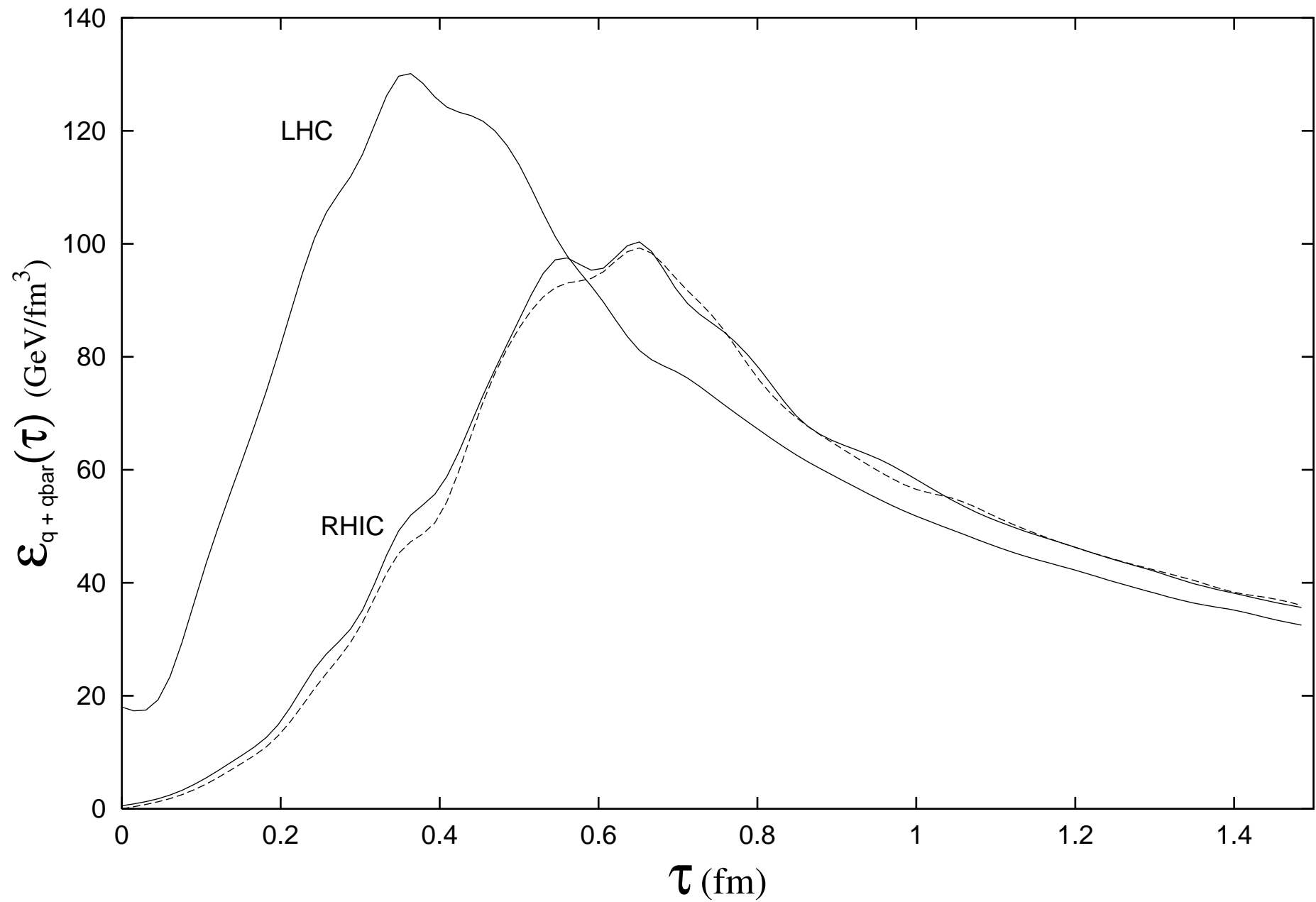
FIG. 7. Energy density for quarks and antiquarks vs proper time. The
dashed line as in Fig. 1.

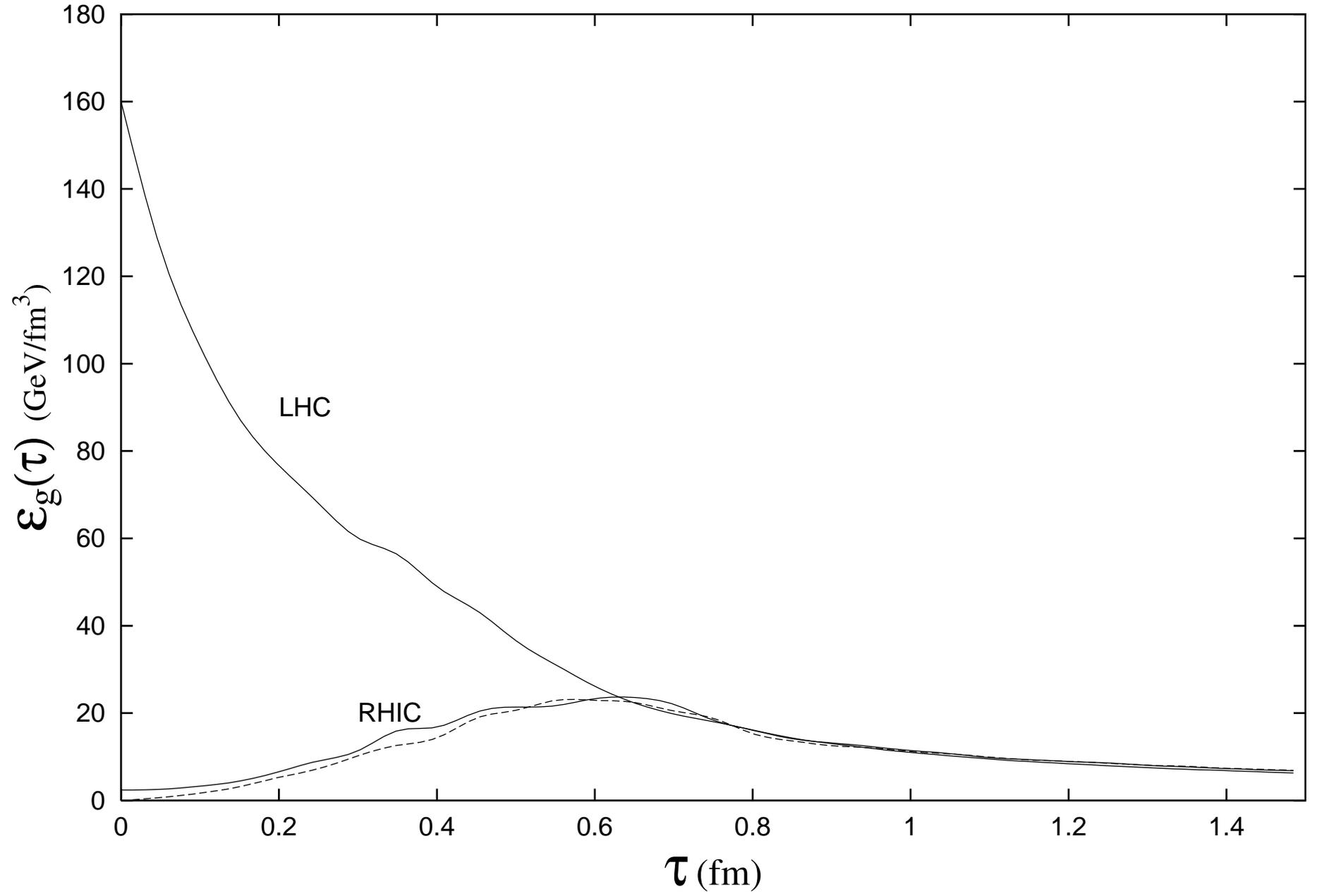
FIG. 8. Energy density for gluons vs proper time. The dashed line as
in Fig. 1.


\begin{thebibliography}{99}

\bibitem{LGT}
A. Polyakov, Phys. Lett. B {\bf 72}, 477 (1978); 
L. Susskind, Phys. Rev. D {\bf 20}, 2610 (1979);
C. Borgs and E. Seiler, Nucl. Phys. B {\bf 215} [FS7], 125 (1983); 
Comm. Math. Phys. {\bf 91}, 329 (1983).
For the status of the lattice QCD, see Proc. of the 17th Intl. Sympo. on
Lattice Field Theory (Lattice'99), Pisa, Italy, June-July 1999, to appear
in Nucl. Phys. B (Proc. Suppl.).

\bibitem{qm99} For the status of the various signatures, see the Proc.
of the 14th Intl. Conf. on Ultrarelativistic Nucleus-Nucleus
Collisions (QM'99), Torino, Italy, May 1999, to appear in
Nucl. Phys. A.

\bibitem{Bj83} J.D. Bjorken, Phys. Rev. D {\bf 27}, 140 (1983).

\bibitem{gcn1} G.C. Nayak, JHEP {\bf 02} (1998) 005.

\bibitem{gcn2} G.C. Nayak, Phys. Lett. B {\bf 442}, 427 (1998).

\bibitem{xu} X.-M. Xu, D. Kharzeev, H. Satz and X.-N. Wang,
Phys. Rev. C {\bf 53}, 3051 (1996).

\bibitem{low} F.E. Low, Phys. Rev. D {\bf 12}, 163 (1975).

\bibitem{nussinov} S. Nussinov, Phys. Rev. Lett. {\bf 34}, 1286
(1975).

\bibitem{sch} J. Schwinger, Phys. Rev. {\bf 82}, 664 (1951).

\bibitem{casher} A. Casher, H. Neuberger and S. Nussinov,  
Phys. Rev. D {\bf 20}, 179 (1979).

\bibitem{gyulassy} M. Gyulassy and A. Iwazaki, Phys. Lett. B {\bf
165}, 157 (1985).

\bibitem{wong} S.K. Wong, Nuovo Cimento A {\bf 65}, 689 (1970).

\bibitem{nayak1} G.C. Nayak and V. Ravishankar, Phys. Rev. D {\bf 55},
6877 (1997).

\bibitem{nayak2} G.C. Nayak and V. Ravishankar, Phys. Rev. C {\bf 58},
356 (1998).

\bibitem{eskola} K.J. Eskola and M. Gyulassy, Phys. Rev. C {\bf 47},
2329 (1993).

\bibitem{kajantie} K.J. Eskola, K. Kajantie and J. Lindfors,
Nucl. Phys. B {\bf 323}, 37 (1989); K.J. Eskola and K. Kajantie,
Z. Phys. C {\bf 75}, 515 (1997); X.-N. Wang, Phys. Rev. D {\bf 43},
104 (1991).

\bibitem{bbr} K. Kajantie and T. Matsui, Phys. Lett. B {\bf 164},
373 (1985); B. Banerjee, R.S. Bhalerao and V. Ravishankar,
Phys. Lett. B {\bf 224}, 16 (1989); A. Bialas, W. Czyz, A. Dyrek and
W. Florkowski, Nucl. Phys. B {\bf 296}, 611 (1988).

\bibitem{heinz} H.-T. Elze and U. Heinz, Phys. Rep. {\bf183}, 81
(1989).

\bibitem{eskola1} K.J. Eskola, V.J. Kolhinen and P.V. Ruuskanen,
Nucl. Phys. B {\bf 535}, 351 (1998).

\bibitem{eskola2} K.J. Eskola, V.J. Kolhinen and C.A. Salgado,
e-print hep-ph/9807297.

\bibitem{kg95} K. Geiger, Phys. Rep. {\bf 258}, 237 (1995).

\bibitem{xnw97} X.-N. Wang, Phys. Rep. {\bf 280}, 287 (1997).

\bibitem{lm94} L. McLerran and R. Venugopalan, Phys. Rev. D {\bf 49},
2233 (1994); D {\bf 49}, 3352 (1994).

\bibitem{bj75} J.D. Bjorken in Current Induced Reactions: Proc. of the
7th Intl. Summer Inst. on Theoretical Particle Physics, Hamburg 1975,
edited by J.G. K\"orner, G. Kramer and D. Schildknecht,
Springer-Verlag, Berlin, 1976, (Lecture Notes in Physics, Vol. 56),
page 93;
J.D. Bjorken in Proc. of the Summer Inst. on Particle Physics,
Stanford, July 1973, edited by D.W.G.S. Leith and S.D. Drell,
Vol. 1, page 1.

\bibitem{lm99} L. McLerran, private communication.

\bibitem{kv99} A. Krasnitz and R. Venugopalan, Nucl. Phys. B {\bf
557}, 237 (1999).

\bibitem{wp99} W. Poeschl and B. Mueller Phys. Rev. D {\bf 60}, 114505
(1999); S.A. Bass, B. Mueller and W. Poeschl, J. Phys. G {\bf 25},
L109 (1999); A. Kovner, L. McLerran and H. Weigert, Phys. Rev. 
D {\bf 52}, 3809 (1995); D {\bf 52}, 6231 (1995).

\bibitem{bhal} R.S. Bhalerao and V. Ravishankar, Phys. Lett. B {\bf
409}, 38 (1997).

\bibitem{jcrb99} J.C.R. Bloch, C.D. Roberts and S.M. Schmidt, e-print
nucl-th/9910073.

\end{thebibliography}
\end{document}